\documentclass[aps,prl,twocolumn,showpacs]{revtex4-1}
\usepackage{amsfonts}
\usepackage{amsmath}
\usepackage{amsthm}
\usepackage{graphics}
\usepackage{graphicx}
\usepackage{subfigure}
\usepackage{amssymb}
\usepackage{graphics}
\usepackage{graphicx}
\usepackage{epstopdf}
\usepackage{color}
\usepackage{subfigure}
\usepackage{pgfplots,tikz}
\usepackage{url}
\usepackage[toc,page]{appendix}
\newcommand\bet{{g}}
 
\newcommand\alps{{\frac{\hbar^2}{2m}}}

\newcommand\dertt[1]{ \frac{\partial{ #1}}{\partial t} }
\newcommand\gd{\mbox{${\bf \nabla}^{2}$}}

\begin{document}

\title{Evolution of a superfluid vortex filament tangle driven by the Gross-Pitaevskii equation}
\author{Alberto Villois}
\affiliation{School of Mathematics, University of East Anglia, Norwich Research Park, Norwich, NR4 7TJ, United Kingdom}
\author{Davide Proment}
\affiliation{School of Mathematics, University of East Anglia, Norwich Research Park, Norwich, NR4 7TJ, United Kingdom}
\author{Giorgio Krstulovic}
\affiliation{
Laboratoire J.L. Lagrange, UMR7293, Universit\'e de la C\^ote d'Azur, CNRS, Observatoire de la C{\^o}te d'Azur, B.P. 4229, 06304 Nice Cedex 4, France}

%\pacs{67.85.De, 67.25.dk, 47.37.+q, 67.25.dt, 03.75.Kk }

\begin{abstract}
The development and decay of a turbulent vortex tangle driven by the Gross-Pitaevskii equation is studied. 
Using a recently-developed accurate and robust tracking algorithm, all quantised vortices are extracted from the fields. 
The Vinen's decay law for the total vortex length with a coefficient that is in quantitative agreement with the values measured in Helium II is observed. 
The topology of the tangle is then studied showing that linked rings may appear during the decay. 
The tracking also allows for determining the statistics of small-scales quantities of vortex lines, exhibiting  large fluctuations of curvature and torsion. 
Finally, the temporal evolution of the Kelvin wave spectrum is obtained providing evidence of the development of a weak-wave turbulence cascade. 
\end{abstract}
\maketitle

The full understanding of turbulence in a fluid is one of the oldest yet still unsolved problems in physics.
A fluid is said to be turbulent when it manifests excitations occurring at several length-scales.
Due to the large number of degrees of freedom and the nonlinearity of the governing equations of motion, the problem is usually tackled statistically by introducing assumptions and closures in terms of correlators.
This is the case in the seminal work of Kolmogorov in 1941 based on the idea of Richardson's energy cascade, where energy in classical fluids is transferred from large to small scales \cite{frisch1995t}. 

Superfluids form a particular class among fluids characterised essentially by two main ingredients: the lack of dissipation and the evidence that vortex circulation takes only discrete values multiple of the quantum of circulation \cite{donnelly1991}.
Superfluid examples which are routinely created in laboratories are superfluid liquid Helium (He II) and Bose-Einstein condensates (BECs) made of dilute Alkali gases.
Here the superfluid phase is usually modelled via a complex field describing the order parameter of the system and quantised vortices appear as topological defects where the superfluid density vanishes. 

In three spatial dimensions those defects organise themselves into closed lines (or even open lines at the boundaries if confining sides are considered)  of different configurations.
Any vortex line point induces a velocity field in the superfluid which affects the motion of any object in the system including the vortex line itself.
In general, even for a single closed vortex line, the dynamics are chaotic and the problem does not have analytical solutions.  
Superfluid turbulence regards the study of the evolution of many vortex lines, a tangle, which induce velocity field gradients in the fluid at several length scales.

Different mathematical models have been devised to mimic the dynamics of a superfluid. 
An example is the vortex filament (VF) model based on the Biot-Savart law that relates vorticity and velocity \cite{Schwarz_VF}.
This model is able to mimic the dynamics of dense vortex tangles due to a relatively fast numerical integration technique \cite{Baggaley2011}.
The VF model implicitly assumes that the superfluid density field is constant everywhere and the vortex structure is precisely a line with vanishing core.
This assumption is generally satisfied in He II where the characteristic experimental setup sizes, and consequently the largest scales of the motion, are order of $ 10^{-1} $m and the vortex core is order of $ 1 \mathring{A} = 10^{-10} $m. 
Moreover, since Helium is in its liquid phase, the compressibility effects can be usually neglected.
However, the VF model fails to describe vortex reconnections. These are rapid changes in the topology of the vortex configuration which occur naturally in a superfluid \cite{bewley2008characterization} and are one of the main mechanisms responsible for the energy transfer. Reconnections are thus introduced by some ad-hoc mechanism.

Another superfluid model that admits quantised vortices and inherently possesses vortex reconnections is the Gross-Pitaevskii (GP) equation that describes the evolution of the superfluid order parameter $\psi$.
In contrast to the VF model, the GP equation allows density fluctuations in terms of phonons and density depletion at the vortex cores. 
Although it has been formally derived as a mean-field theory for a dilute boson gas in the limit of zero temperature \cite{pitaevskii2003bose}, it also qualitatively reproduces He II dynamics. 
The vortex core size in the GP equation is order of the healing length $ \xi $, the only intrinsic characteristic length-scale of the model; nowadays experimental techniques are able to create BEC setups that are $ 10^1-10^2 $ healing lengths where superfluid turbulence can develop \cite{Henn2009}. 
In turbulent superfluids, vortices constantly re-arrange themselves following reconnections into complex tangles with non-trivial geometrical, algebraic and topological properties \cite{Barenghi2001197}. 
At small scales, helical excitations of vortex lines known as Kelvin waves (KWs), are believed to be the ultimate mechanism of energy dissipation via phonon emission \cite{PhysRevB.64.134520}. 
To study such dynamics, the GP equation has the advantage that no extra modelling is needed (unlike the VF model). 
However GP does not provide direct information on vortices.

In this Letter we apply a novel numerical algorithm \cite{2016arXiv?} to track accurately the configuration of a turbulent vortex tangle evolving accordingly to the GP model. 
We focus on the evolution and decay of the tangle.
Firstly, we show that after the onset of turbulence, the vortex line density satisfies the Vinen's decay law \cite{VinenDecay} with a coefficient that is in agreement with He II. 
Different algebraic and topological quantities of the tangle are then measured. 
The tracking allows for obtaining curvature and torsion distributions of the vortex tangle. 
Finally, we perform a direct measurement of KWs during the dynamics and compute a KW spectrum that appears to be consistent with the L'vov-Nazarenko theoretical prediction \cite{springerlink:10.1134}.
    
The GP model for the condensate wave-function $\psi$ is given by
\begin{equation}
i\hbar\dertt{\psi} =- \alps \gd \psi + \bet|\psi|^2\psi ,
\label{Eq:GPE}
\end{equation}
where $m$ is the mass of the bosons and $g=4 \pi a \hbar^2 / m$, with $a$ the $s$-wave scattering length. 
Madelung's transformation $\psi({\bf x},t)=\sqrt{\rho({\bf x},t)/m}\exp{[i \frac{m}{\hbar}\phi({\bf x},t)]}$ relates the wave-function $\psi$ to a superfluid of density $\rho({\bf x},t)$ and velocity ${\bf v}={\bf \nabla} \phi$. 
The quantum of circulation about the $ \psi = 0 $ vortex lines is $ \Gamma=h/m $. When Eq.\eqref{Eq:GPE} is linearised about a constant value $\psi= \hat{\psi}_{\bf 0}$, the sound velocity is given by $c={(g| \hat{\psi}_{\bf 0}|^2/m)}^{1/2}$ with dispersive effects taking place at length scales smaller than the healing length $\xi={(\hbar^2/2m|\hat{\psi}_{\bf 0}|^2g) }^{1/2}$.    
In the simulations presented in this Letter, the mean density is fixed to the unity and the physical constants in Eq.\eqref{Eq:GPE} are determined by the values of $\xi$ and $c=1$. The quantum of circulation results in $\Gamma=4\pi c\,\xi/\sqrt{2}$. 
Numerical integration of Eq.\eqref{Eq:GPE} is performed using a standard pseudo-spectral code. We integrate the so-called Taylor-Green flow \cite{nore1997decaying} with no enforced symmetries at resolutions $256^3$ and $512^3$ with $\xi=2\pi/256$ and $\xi=2\pi/512$ respectively (see Appendix B). 
With these units the largest eddy turnover time is order of the unity.

\begin{figure}
\includegraphics[width=0.45\textwidth]{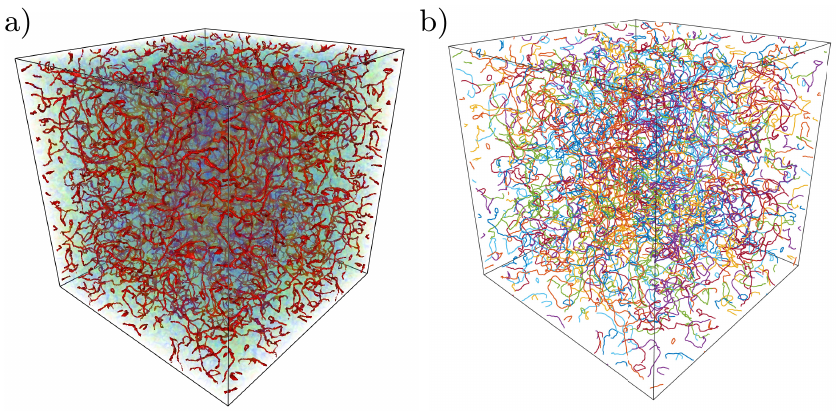}
\caption{(Color online) a) Isosurface of density field at $t=12$ rendered with VAPOR software. Low-density regions corresponding to vortex lines are plotted in red, while density fluctuations (sound) are rendered in light blue. b) Corresponding vortex tracking. Different colors correspond to different vortices. Resolution $256^3$.}\label{Fig:Snapshot}
\end{figure}

The Taylor-Green flow initially contains a configuration of unstable large-scale rings that develop to create a turbulent tangle. 
Vortices can be easily spotted by plotting the low-value iso-surfaces of the density field as displayed in Fig.\ref{Fig:Snapshot}a. Low-density regions corresponding to vortex lines are plotted in red, while density fluctuations (sound) are rendered in light blue.
We track the vortex lines forming the tangle with a recently-developed algorithm \cite{2016arXiv?}.
Vortex lines are followed using the pseudo-vorticity field \cite{2014arXiv1410.1259R} and the exact vortex position is obtained by a Newton-Raphson method.
The algorithm is robust and accurate as it takes full advantage of the spectral resolution of the field. 
It allows for identifying separately each single line forming the tangle. 
Figure \ref{Fig:Snapshot}b shows the corresponding tracked tangle displaying in different colors all the 496 vortex rings.

\begin{figure}
\includegraphics[width=0.49\textwidth]{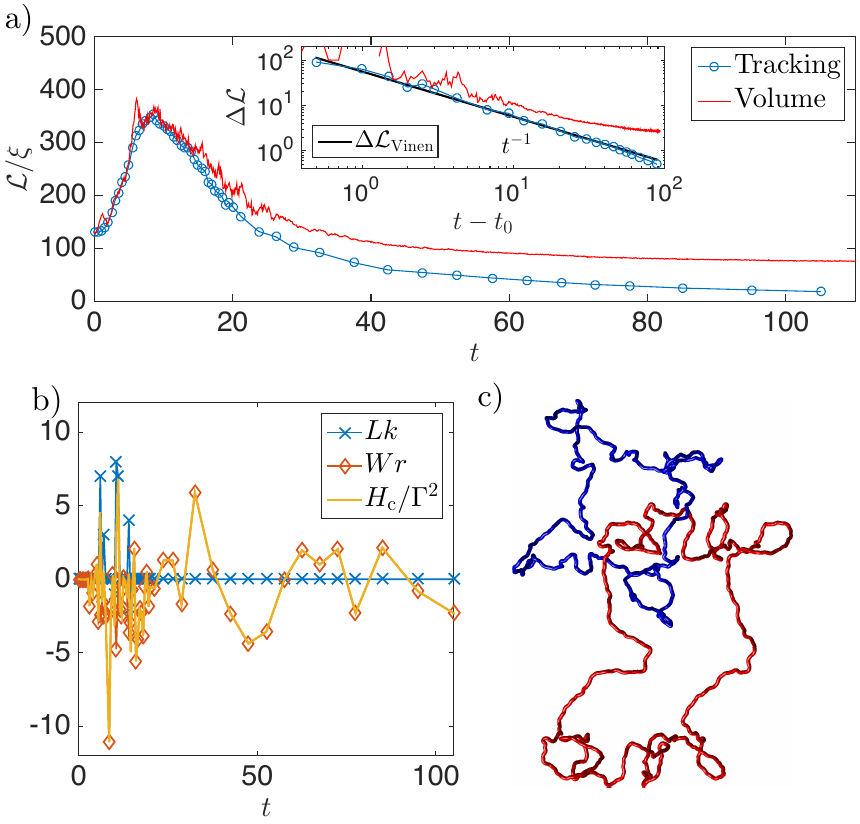}
\caption{(Color online) a) Temporal evolution of the vortex line density: tracked data are plotted using blue circles, volume estimation in solid red line. The inset shows the long time decay of  $\Delta \mathcal{L}$ (see text), together with Vinen's prediction $ \Delta \mathcal{L}_{\rm Vinen}=[\chi_2\frac{\Gamma}{2 \pi}(t-t_0)]^{-1}$, with $\chi=0.65$ (solid black line). b) Temporal evolution the total linking $ Lk $, writhe $ Wr $ and center-line helicity $ H_c $. c) Visualisation of two linked rings at  $t=21$. Resolution $256^3$.}
\label{Fig:EvolutionLineQuantities}
\end{figure}

During the decay, vortices radiate phonons that populate the small-scales creating a thermal bath that exchanges energy and momentum with the vortices. 
This process mimics mutual friction and leads eventually to the total annihilation of vortex rings  \cite{Krstulovic2011b}. 
In superfluids such a decay is modelled by the Vinen equation \cite{VinenDecay} for vortex line density $\mathcal{L}$:
\begin{equation}
\frac{d \mathcal{L}}{d t} = -\chi_2 \frac{\Gamma}{2\pi} \mathcal {L}^2 \, , 
\label{eq:Vinen}
\end{equation} 
where $ \chi_2$ is a constant of order of unity. 
Its solutions manifest a $ \mathcal{L} \sim t^{-1} $ behaviour at long times: this power-law decay has been named quantum turbulent decay and measured in He II experiments \cite{PhysRevLett.100.245301} and BS numerical simulations \cite{PhysRevB.85.060501}.  
In Fig.\ref{Fig:EvolutionLineQuantities}a we show the temporal evolution of $\mathcal{L}$.  
It is worth noticing that it grows at the initial stages: this is caused by the instability of the initial Taylor-Green configuration and the subsequent vortex stretching due to numerous vortex reconnections. 
The data is compared with an estimation of $ \mathcal{L} $ obtained by computing the ratio between the volume of points such that $\rho({\bf x})<0.2$ and the corresponding surface of a perfect two-dimensional vortex profile. 
This latter method is a standard technique within GP to estimate $\mathcal{L}$ numerically \cite{PhysRevLett.104.075301}.
Even if this technique is able to capture the qualitative behaviour of $ \mathcal{L} $, it fails to grasp at long times the power-law predicted by Vinen's equation. 
This is show in the inset of  Fig.\ref{Fig:EvolutionLineQuantities}a where the measured $\Delta\mathcal{L}(t)=(\mathcal{L}(t)^{-1}-\mathcal{L}(t_0)^{-1})^{-1}$, setting $t_0=17$, is compared to Vinen's prediction.
This discrepancy can be explained by the presence of sound waves and small scale KWs not detected by the estimation. The tracking data also allow for determining the numerical constant $\chi_2=0.65$. This value is in remarkable agreement with experimental values measured in He II in the low temperature limit \cite{VinenDecay}.

From Fig.\ref{fig:Tangle} in Appendix A, it is clear that the complexity of tangle first increases and then decreases. 
This observation is intuitively explained by the idea that vortex lines simultaneously stretch, bend and coil during reconnection events. 
The complexity of tangle can be measured by computing the changes in some of its algebraic and topological quantities, such as the total number of rings, the average crossing number, the total linking $ Lk $ and writhe $ Wr $ \cite{Barenghi2001197}. 
Explicit definitions are given in Appendix C. 
The total number of rings and the average crossing number vary, as expected, in a similar manner to the total vortex length in Fig.\ref{Fig:EvolutionLineQuantities}a (See Fig.3 in the Appendix C). 
Here we focus on the central line helicity $ H_c /\Gamma^2 = Lk+Wr $ \cite{Scheeler28102014}. 
This quantity is the analogous to helicity in classical fluid dynamics, an important inviscid invariant. 
The linking number $Lk$ takes integer values and gives information about the number of linked rings present in the system, whereas the writhe takes real values and its contribution comes from knots and KWs \cite{diLeoniHelicity}. 
Figure \ref{Fig:EvolutionLineQuantities}b shows the temporal evolution of these three quantities. Initially, $Lk=Wr=0$, as expected for the Taylor-Green flow. 
Surprisingly, during the evolution $Lk$ becomes non-zero, indicating the presence of linked rings, such as the ones displayed in Fig.\ref{Fig:EvolutionLineQuantities}c \footnote{Although linked rings are present in the flow, the probability of finding them is very small. It can be estimated counting all the linked rings and it is at most of order $10^{-4}$ for the Taylor-Green flow.}. 
This is remarkable as in the GP model sufficiently simple vortex configurations usually decay by reducing their linking number \cite{Kleckner:2016yu}. 
Once the decay is established, no linked rings are present and only writhe contributes to $H_c$. The center-line helicity fluctuates about a zero mean, an indication of the presence of KWs in this turbulent tangle, as indirectly observed in \cite{diLeoniKW}.

\begin{figure}
\includegraphics[width=0.49\textwidth]{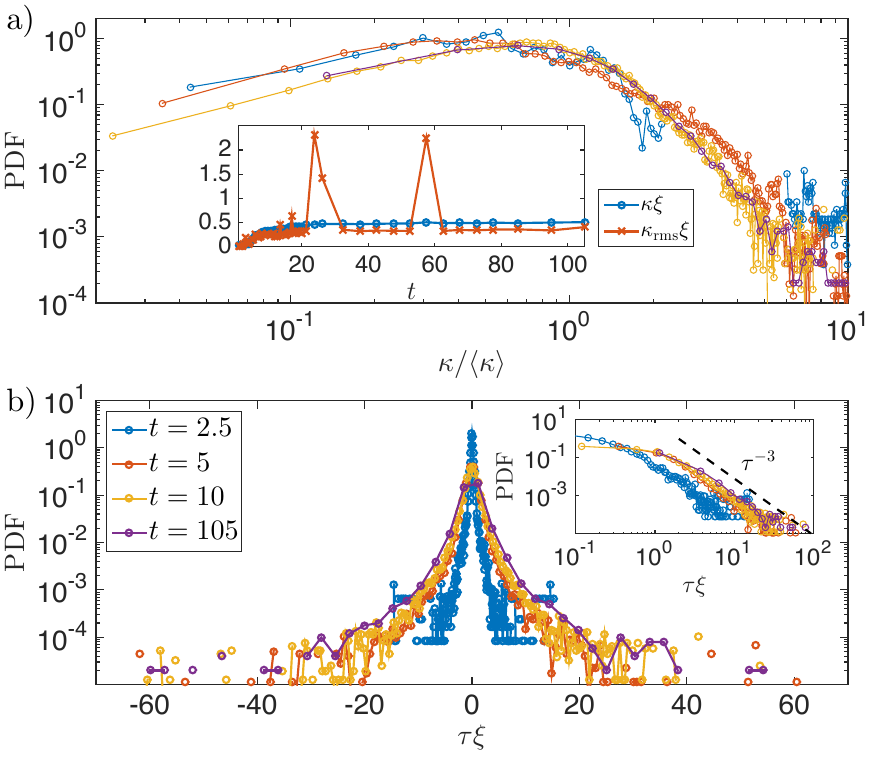}
\caption{(Color online) a) PDFs of curvature $\kappa$ normalised by their respective mean values $\langle \kappa\rangle$ at different times (same legend as (b)). The inset displays the temporal evolution of the mean and rms values of $\kappa$.
b) PDFs of torsion $\tau$ at different times. The inset emphasises their $\tau^{-3}$ power-law tail. Resolution $256^3$.}
\label{Fig:Statistics}
\end{figure}

We now study statistical properties of some geometrical quantities of the vortex filaments. We explore the time behaviour of the probability density functions (PDFs) of the curvature $ \kappa $ and torsion $ \tau $.
In Fig.\ref{Fig:Statistics}a we present the PDF of curvature, normalised by its mean value, at different stages.
The temporal evolution of the mean curvature  $ \langle\kappa\rangle$ and its \emph{rms }value $\kappa_{\rm rms}$ are also displayed in the inset.
We can observe that  $ \langle\kappa\rangle$ increases rapidly at early stages and then almost saturates. The rms value of the curvature presents the same tendency with the exception of peaks. These are evidences of reconnection events where high values of curvature are found in localised regions. 
It is worth noticing that the PDFs, rescaled by their mean curvature, exhibit a relatively good collapse to a self-similar form.
This latter observation indicates a power-law behaviour $\sim \kappa^{1} $ at small curvature values, while an exponentially-decaying tail is present at large curvature values. 
A similar behaviour has also been observed within the VF model \cite{KondaurovaVFMStat}.
In Fig.\ref{Fig:Statistics}c we plot the torsion PDFs at the same stages.
The mean torsion is always about zero and there is no evidence of any skewness in the PDFs.
The distributions' tails show an universal power-law behaviour of $ \tau^{-3} $ at all times, meaning that the second and higher moments of the torsion diverge during the decay.
The same scaling appears in vortex tangles of random wave fields that are solutions of the Helmholtz equation \cite{1751-8121-47-46-465101}.
This might be an indication that for one-time small-scale quantities, quantum turbulent tangles could be interpreted as random vortices.

\begin{figure}
\includegraphics[width=0.45\textwidth]{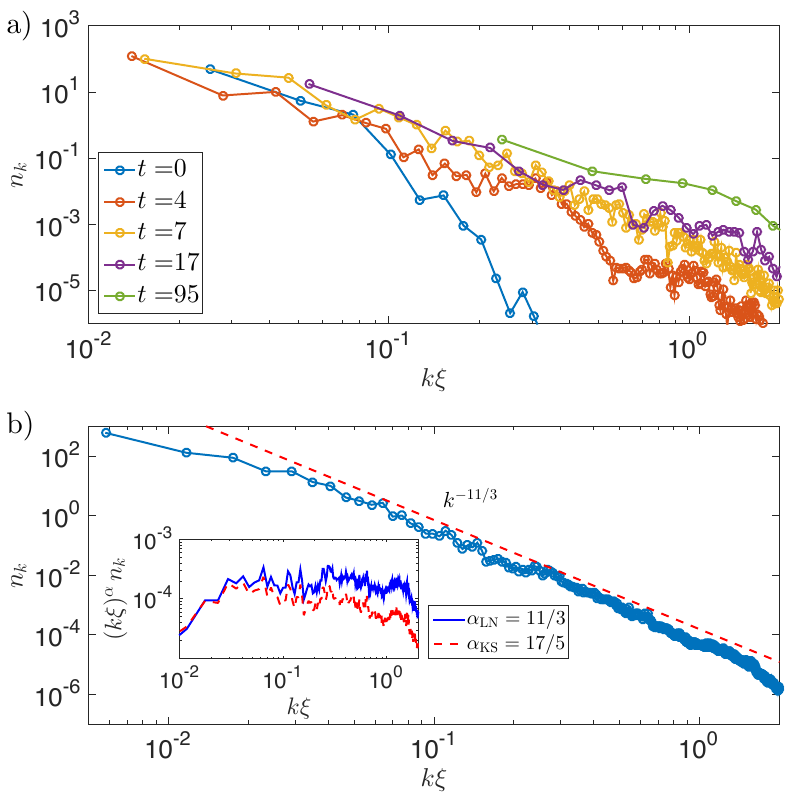}
\caption{(Color online) a) Temporal evolution of KW spectra (averaged over the 50 longest rings). Resolution $256^3$. b) KW spectrum at $t \sim 5$ (averaged over the 50 longest rings) for run at resolution $512^3$. The dashed line displays the $k^{-11/3}$ scaling. The inset displays the respective $k^{11/3}$ (solid blue) and $k^{17/5}$ (dashed red) compensated spectra.}
\label{Fig:KWspectra}
\end{figure}

The large curvature fluctuations and the torsion fluctuation about a zero mean suggest also the presence of KWs at all scales propagating on quasi-planar vortex rings.   
By exploiting the accuracy of the tracking algorithm we are able to directly detect KWs on those rings. Competing theories have been put forward to statistically predict a power-law KW spectrum in the form of $ n_k \sim k^{-\alpha} $ (here $ k $ is the Kelvin wavenumber) and to explain the energy transfer through KW scales. Assuming small amplitude KWs (weak non-linearity), Kozik\&Svistunov \cite{PhysRevLett.92.035301} and L'vov\&Nazarenko \cite{springerlink:10.1134} obtained the exponents $ \alpha_{KS}=17/5 $ and $ \alpha_{LN}=11/3 $ respectively considering two different orders of interaction.
We compute the KW spectra of the $50$ largest rings during the evolution of the tangle applying a Gaussian kernel in order to establish the unperturbed ring (see Appendix D).
The spectra, averaged over the rings, are shown for different times in Fig.\ref{Fig:KWspectra}a.    
It is evident that all accessible KW modes get populated at early times due to reconnection events that trigger the cascade \cite{PhysRevLett.86.3080}.
We observe KW spectra exhibiting power-laws with an exponent independent of time where the best scaling is appreciated at the time where the rings are the longest ($4 \le t \le 7$).
To get the best estimation of the power-law exponent, we repeated the Taylor-Green decay in a simulation box twice larger; in this new configuration the scaling range spans almost two wavenumber decades.
In Fig.\ref{Fig:KWspectra}b we show the spectrum at $ t \sim 5 $: the observed power-law exponent agrees with the L'vov\&Nazarenko  $ \alpha_{LN}=11/3 $ prediction.
This can be better appreciated by looking at the compensated spectra with respect to $ \alpha_{LN} $ and $ \alpha_{KS} $ showed in the inset.
This finding corroborates the result in favour of L'vov\&Nazarenko's prediction previously obtained while studying the KW oscillations about a perfect straight line in the GP model \cite{PhysRevE.86.055301}. 
We highlight that although the weak-wave turbulence prediction for the KW spectrum was formally derived for KWs on an isolated straight vortex line using the VF model, it remarkably turns out to be valid in a dense turbulent tangle also driven by the GP model. 
This is certainly due to the fact that the predicted KW spectrum was found for the longest rings. 
Small rings quickly loose their energy by phonon radiation and also exchange momentum with sound waves. Both contributions are important to understand dissipation of superfluids at very low temperature and further studies are still needed to fully comprehend the relevance of such mechanisms.

Tracking vortices in GP turbulence opens a new way for studying and understanding the topological configuration and properties of quantum vortex tangles. 
Although unlikely, we show that rings can link creating a local (in time and space) fluctuation of helicity. 
It will be of great interest to repeat a similar analysis in a GP setting where the mean helicity of the flow is not zero, like the ABC flow introduced in \cite{diLeoniHelicity} where linking and self-linking processes could be substantially enhanced. 
Overall, the results presented in this Letter confirm that some predictions traditionally associated to superfluid liquid Helium become important in weakly-interacting BECs at low temperature described by the GP model. Nowadays BEC experimentalists are able to create and track few vortices in harmonic traps \cite{Lamporesi2013VortexCreation,Serafini2015_VortexBEC}. A controlled experimental setting with a turbulent BEC, such as the one presented in this Letter, has yet to be achieved but it should be realisable in the near future.

%\newpage
\begin{acknowledgments}  
GK, DP and AV were supported by the cost-share Royal Society International Exchanges Scheme (ref. IE150527) in conjunction with CNRS.
Computations were carried out on the M\'esocentre SIGAMM hosted at the Observatoire de la C\^ote d'Azur and on the High Performance Computing Cluster supported by the Research and Specialist Computing Support service at the University of East Anglia.
\end{acknowledgments}

\appendix
\section{Appendix A. Vortex Tracking}\label{App.A}
We have recently developed a robust and accurate algorithm to track vortex lines in the Gross-Pitaevskii equation (GP) with arbitrary geometries in a periodic domain. 
The full details of the algorithm and the case studies to check its validity can be found in \cite{2016arXiv?}.
We recall here briefly the basic ideas.
A quantised vortex line is defined by the nodal lines of the wavefunction. In three dimensions this corresponds to a line defined by 
\begin{equation}
Re[\psi(x,y,z)]=Im[\psi(x,y,z)]=0
\end{equation}
The algorithm is based on a Newton-Raphson method to find zeros of $\psi$ and on the knowledge of the \emph{pseudo-vorticity} field ${\bf W}=\nabla Re[\psi]\times \nabla Im[\psi]$, always tangent to the line, to follow vortex lines (in the spirit of Rorai et al. \cite{2014arXiv1410.1259R}). 

\begin{figure}[h]
 \centering
      \includegraphics[scale=0.35]{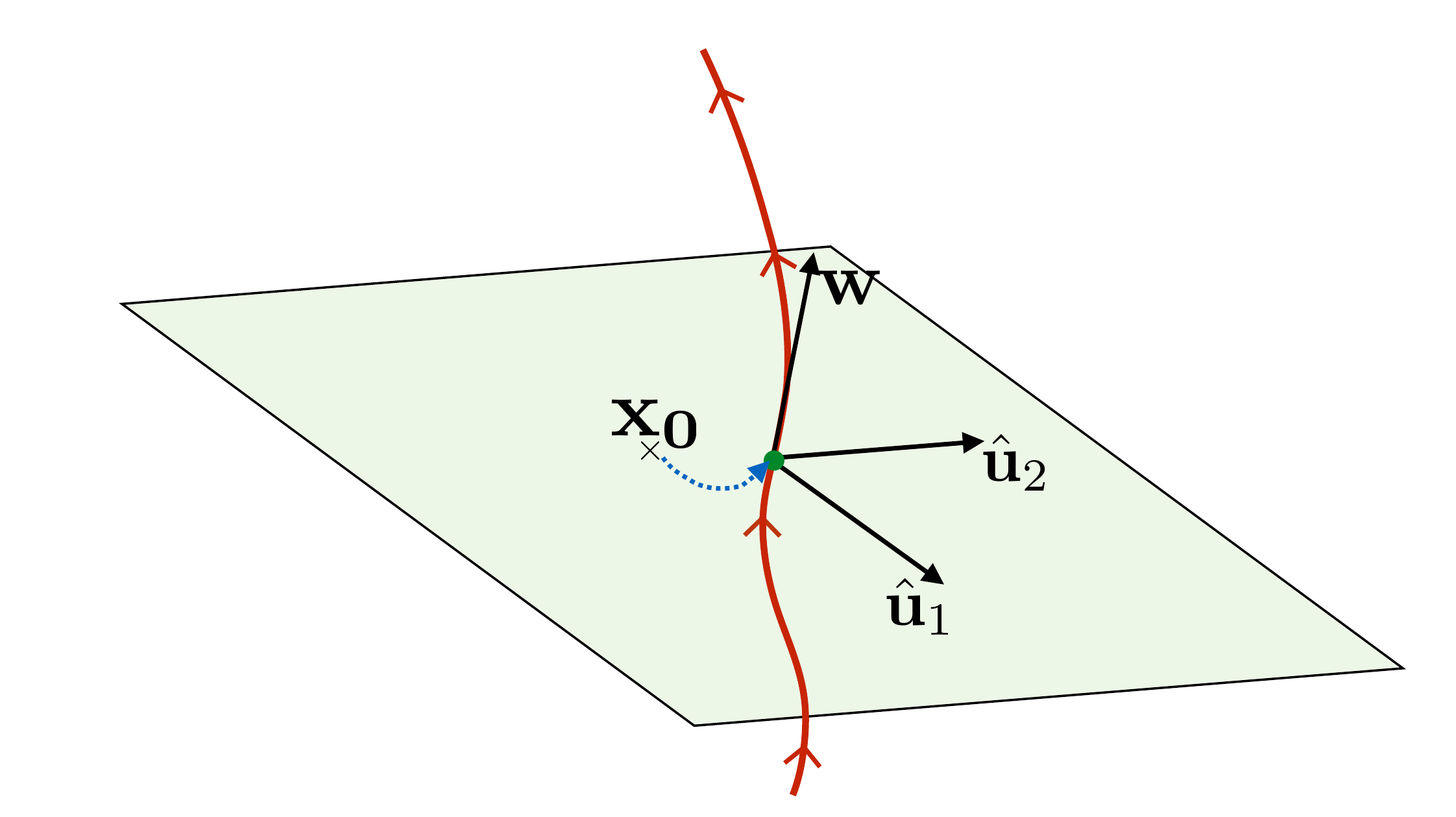}
 \caption{Sketch of the plane on which the Newton-Raphson method is implemented.
\label{fig:Sketch}}
\end{figure}

Starting from a point ${\bf x_0}$ where the density $|\psi|^2$ is below a given small threshold (therefore very close to a vortex), we define the orthogonal plane to the vortex line using ${\bf W}({\bf x_0})$. 
The plane is then spanned by the two directors $\hat{{\bf u}}_1$ and $\hat{{\bf u}}_2$ as illustrated in Fig.\ref{fig:Sketch}.
%In that plane%, the position of the vortex ${\bf x_{\rm v}}$ is found by using a Newton-Raphson method. Starting from ${\bf x_0}$, the next step of the we use a Newton-Raphson method leads to
A better approximation for the true point ${\bf x_{\rm v}}$ where the vortex lies on the plane %${\bf x_{\rm v}}$  
is then given by ${\bf x_1=x_0+\delta x}$. Here the increment ${\bf \delta x}$ is obtained using the Newton-Raphson formula:
\begin{equation}
0=\psi({\bf x_0+\delta x})\approx\psi({\bf x_0})+ J({\bf x_0}) {\bf \delta x},
\end{equation}
where $J({\bf x_0})$ is the Jacobian matrix expressed as
\begin{equation}\label{Jac_gen}
J=\begin{pmatrix}
 \nabla Re[\psi]\cdot \hat{{\bf u}}_1  &   \nabla Re[\psi]\cdot \hat{{\bf u}}_2 \\
 \nabla Im[\psi]\cdot \hat{{\bf u}}_1&  \nabla Im[\psi]\cdot \hat{{\bf u}}_2
  \end{pmatrix}.
\end{equation}
The increment can be therefore found using  ${\bf \delta x}= - J^{-1}({\bf x_0})\cdot  (Re[\psi({\bf x_0})],Im[\psi({\bf x_0})])^T$. The Jacobian matrix is a non-singular $2\times2$ matrix so its inverse is computed with its explicit formula. 
We underline that the method will require in general the evaluation of the Jacobian~\eqref{Jac_gen} at intermesh points. However, making use of the spectral representation of $\psi$, we can compute spatial derivatives of the field $ \psi $ efficiently using fast Fourier transforms. 
This process can be iterated until the true vortex location ${\bf x_{\rm v}}$ is determined upon a selected convergence precision. 

\begin{figure}[h]
 \centering
  \includegraphics[width=0.45\textwidth]{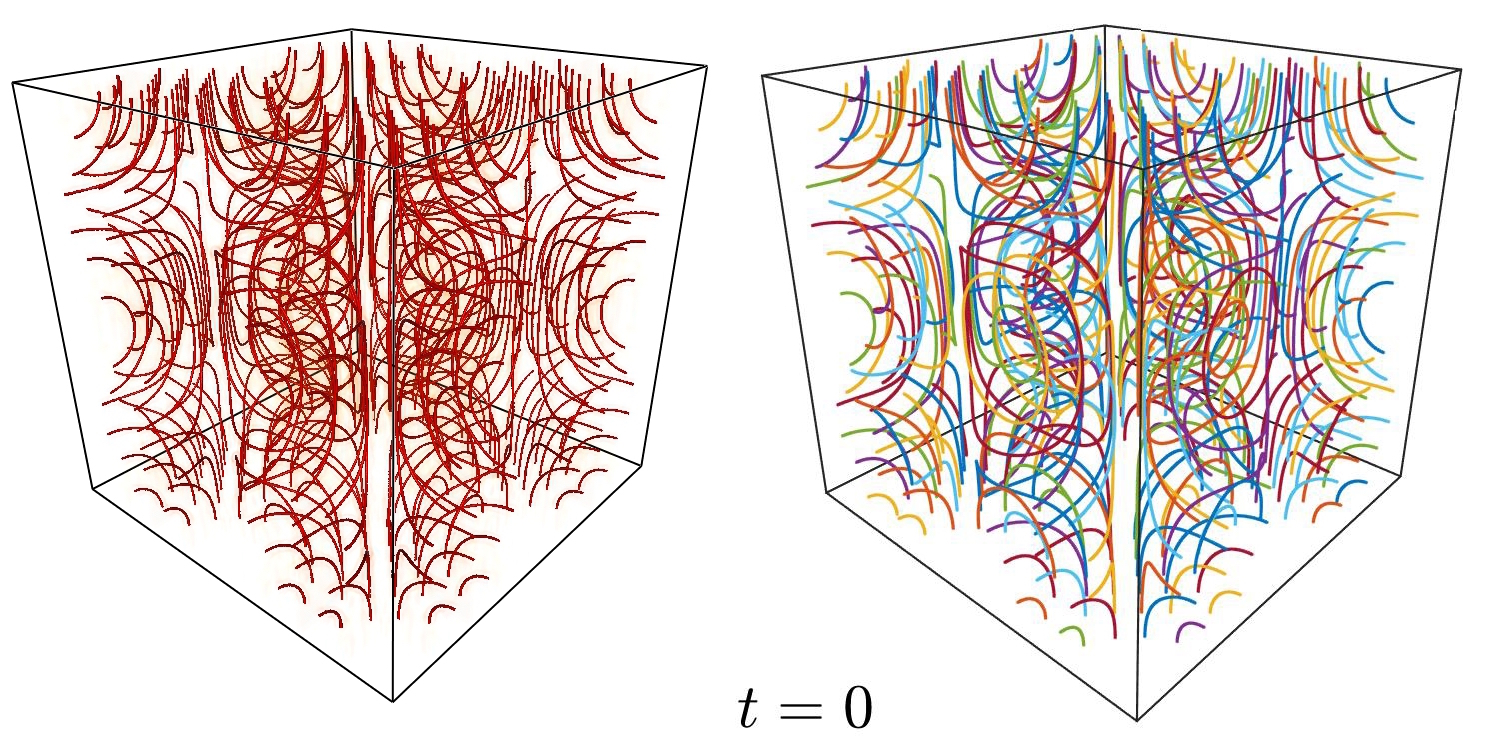}
    \includegraphics[width=0.45\textwidth]{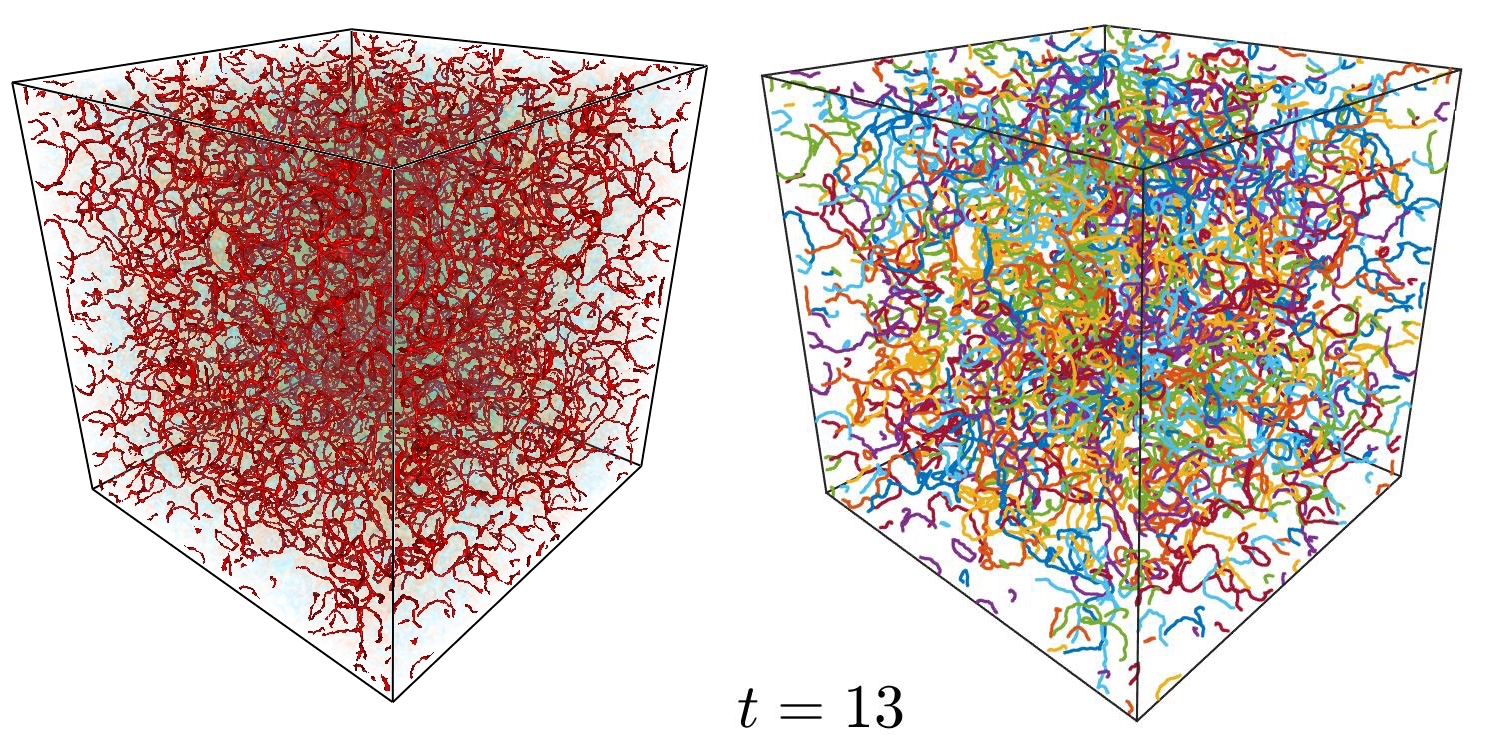}
      \includegraphics[width=0.45\textwidth]{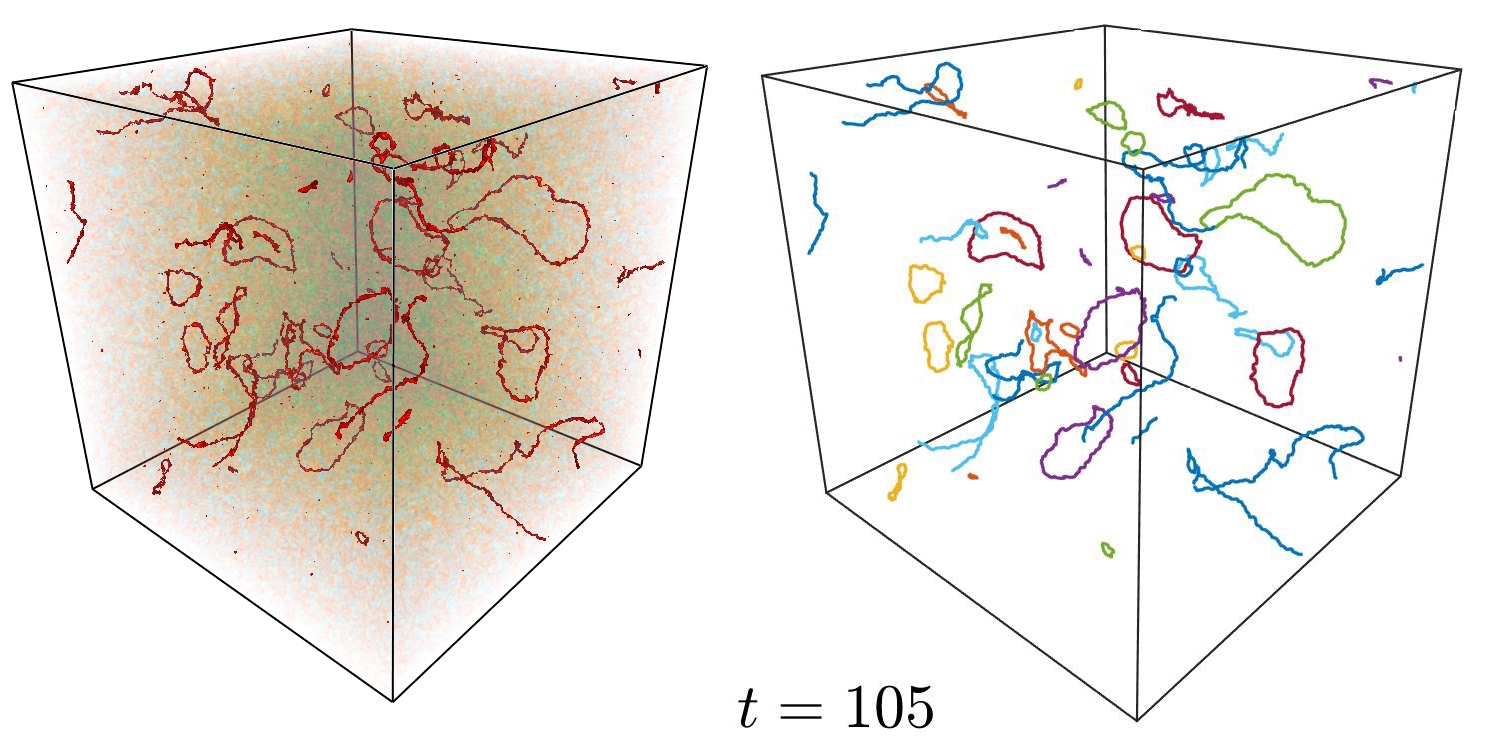}
 \caption{(Color online) (left) Isosurfaces of density field at different times rendered by VAPOR (https://www.vapor.ucar.edu). Low-density regions corresponding to vortex lines are plotted in red, while density fluctuations (sound) are rendered in light blue. (right) Corresponding vortex tracking. Different colours correspond to different vortices. Resolution $256^3$.}
\label{fig:Tangle}
\end{figure}

To track the next vortex point we use as next initial guess ${\bf x_0}={\bf x_{\rm v}}+\zeta {\bf W}$, which is obtained evolving along $\boldsymbol{\omega}_{ps} $  by a small step $\zeta$.
The process is repeated until the entire line is tracked and closed.
Once the line is completely tracked the process is repeated with another line until the whole domain has been fully explored. 
The algorithm has been extensively described and validated in \cite{2016arXiv?} using different case tests. As a final remark, we underline that the algorithm takes full advantage of the spectral resolution when a field is generated integrating the Gross-Pitaevskii with a pseudo-spectral method. A comparison between the low-value isosurfaces of the density field and tracked vortices is displayed in Fig.\ref{fig:Tangle} for different times of the Taylor-Green decay.

\section{Appendix B. Taylor-Green initial condition}
In this work we use the so-called Taylor-Green initial condition introduced by Nore et al. \cite{nore1997decaying}. We recall here for the sake of completeness how this initial condition is produced making use of the Clebsch representation of a vector field. 
Let us assume that an incompressible velocity field ${\bf u_{\rm adv}}(x,y,z)$ admits a global Clebsch representation in terms of the potentials $\lambda(x,y,z)$ and $\mu(x,y,z)$  \cite{lamb1932hydrodynamics}, such that ${\bf u_{\rm adv}}=\lambda\nabla\mu-\nabla \Phi$. It follows that
\begin{equation}
\nabla\times {\bf u_{\rm adv}}=\nabla \lambda\times\nabla \mu.
\end{equation}
The Clebsch representation has an interesting geometrical representation. A vortex line of the velocity flow ${\bf u_{\rm adv}}$ is mapped into a point in the $(\lambda, \mu)$ plane. Indeed a vortex line $\bf{\omega}(s)$ is defined by $\frac{d\bf{\omega}}{ds}=\nabla\times {\bf u_{\rm adv}}(\bf{\omega}(s))$. A simple solution is given by $\lambda(\bf{\omega}(s))=\lambda_{\rm v}$ and $\mu(\bf{\omega}(s))=\mu_{\rm v}$, with $\lambda_{\rm v}$ and $\mu_{\rm v}$ constants.
It follows that, using a $2D$ wavefunction $\psi_{2D}$ of a vortex at the origin, it is straightforward to construct a $3D$ wavefunction with a nodal line defined by $\lambda(x,y,z)=\lambda_{\rm v}$ and $\mu(x,y,z)=\mu_{\rm v}$. This $3D$ field simply reads
\begin{equation}
\psi(x,y,z)=\psi_{2D}(\lambda(x,y,z)-\lambda_{\rm v},\mu(x,y,z)-\mu_{\rm v})
\end{equation}

This idea is applied to the Taylor-Green velocity flow defined in the domain $[0,2\pi]^3$ as:
\begin{equation}
{\bf u_{\rm adv}}=A(\sin{(x)}\cos{(y)}\cos{(z)},- \cos{(x)}\sin{(y)}\cos{(z)},0).
\end{equation}
It is composed of 8 fundamental boxes that are obtained one from each other by mirror symmetric transformation of the sub-box  $[0 ,\pi]^3$. It admits a decomposition in terms of the Clebsch potentials (see \cite{nore1997decaying} for further details):
\begin{equation}
\begin{split}
\lambda(x,y,z)&=\sqrt{A}\cos(x)\sqrt{2 |\cos{(z)}|},\\ \qquad\mu(x,y,z)&=\sqrt{A}\cos(y)\sqrt{2 |\cos{(z)}|}\,{\rm sign}{[\cos{(z)}|},
\end{split}
\end{equation}
in the sense that $\nabla\times {\bf u_{\rm adv}}=\nabla \lambda\times\nabla \mu$. Note that the circulation around the plane $[0,\pi]\times[0,\pi]$ is $8A$

A set of vortex lines can be defined in the wavefunction as
\begin{equation}
\begin{split}
\psi_4(\lambda,\mu)=&\psi_{2D}(\lambda-\frac{1}{\sqrt{2}},\mu)\,\psi_{2D}(\lambda+\frac{1}{\sqrt{2}},\mu)\,\\
&\times\psi_{2D}(\lambda,\mu-\frac{1}{\sqrt{2}})\,\psi_{2D}(\lambda,\mu+\frac{1}{\sqrt{2}})
\end{split}
\end{equation}
Any $2D$ vortex profile approximation can be used as it will be later relaxed by using the Advective-Real-Ginzburg-Landau equation  \cite{nore1997decaying}.
In order to match (as close as possible) the circulation to the Taylor-Green flow, the initial condition is defined as
\begin{equation}
\psi_{\rm TG}(x,y,z)=\psi_4(x,y,z)^{n_c}
\end{equation}
where $n_c=\lfloor (8 A/\Gamma)/4 \rfloor$ and $\Gamma$ represent the quantum of circulation. The wavefunction $\psi_{\rm TG}$ is then evolved under the  Advective-Real-Ginzburg-Landau equation:
\begin{equation}
\hbar\dertt{\psi} = \alps \gd \psi - \bet|\psi|^2\psi +\mu\psi-\hbar\left(i {\bf u_{\rm adv}}\cdot\nabla\psi+\frac{u_{\rm adv}^2}{2\hbar/m}\psi   \right),
\end{equation}
subject to the chemical potential $\mu$. This equation is just the imaginary time evolution of the Gross-Pitaevskii equation after a Galilean transformation with a non-constant velocity field $u_{\rm adv}$. The final states contains a clean (without sound) initial condition with a set of rings (of charge one) distributed along the vortical lines of the Taylor-Green velocity flow. A visualisation of the Taylor-Green flow is displayed in Fig.\ref{fig:Tangle}a. The initial condition is then evolved using the Gross-Pitaevskii equation.

\section{Appendix C. Topological quantities}

The total crossing $\bar{C}$,  total linking $ Lk $, writhe $ Wr $ and center-line helicity $ H_c /\Gamma^2 = Lk+Wr $ were computed directly performing the line integrals over the vortex ring(s) \cite{Barenghi2001197}. They are defined as
\small
\[ \bar{C}=\sum_{i\neq j}=C_{i,j}\qquad C_{i,j} =\frac{1}{4\pi} \oint_{\mathcal{C}_i} \oint_{\mathcal{C}_j} \left| \frac{({\bf R}_i-{\bf R}_j)\cdot \mathrm{d}{\bf R}_i \times \mathrm{d}{\bf R}_j}{|{\bf R}_i-{\bf R}_j)|^3}\right |\]
\[
Lk=\sum_{i\neq j}= Lk_{ij} \qquad Lk_{ij}=\frac{1}{4\pi}  \oint_{\mathcal{C}_i} \oint_{\mathcal{C}_j} \frac{({\bf R}_i-{\bf R}_j)\cdot \mathrm{d}{\bf R}_i \times \mathrm{d}{\bf R}_j}{|{\bf R}_i-{\bf R}_j)|^3}\]
\[
Wr=\sum_{i}Wr_i \qquad Wr_i=\frac{1}{4\pi} \oint_{\mathcal{C}_i} \oint_{\mathcal{C}_i} \frac{({\bf R}_i-{\bf R}_i')\cdot \mathrm{d}{\bf R}_i \times \mathrm{d}{\bf R}_i'}{|{\bf R}_i-{\bf R}_j)|^3} \, ,
\]
\normalsize
where $ \bf{R}_i$ and $ \bf{R}_i' $ correspond the points identifying the $ i $th ring $ \mathcal{C}_i$. For the writhe number, ${\bf R}_i$ and ${\bf R}_i'$  correspond to two different points of the same ring.

\begin{figure}
  \includegraphics[width=0.45\textwidth]{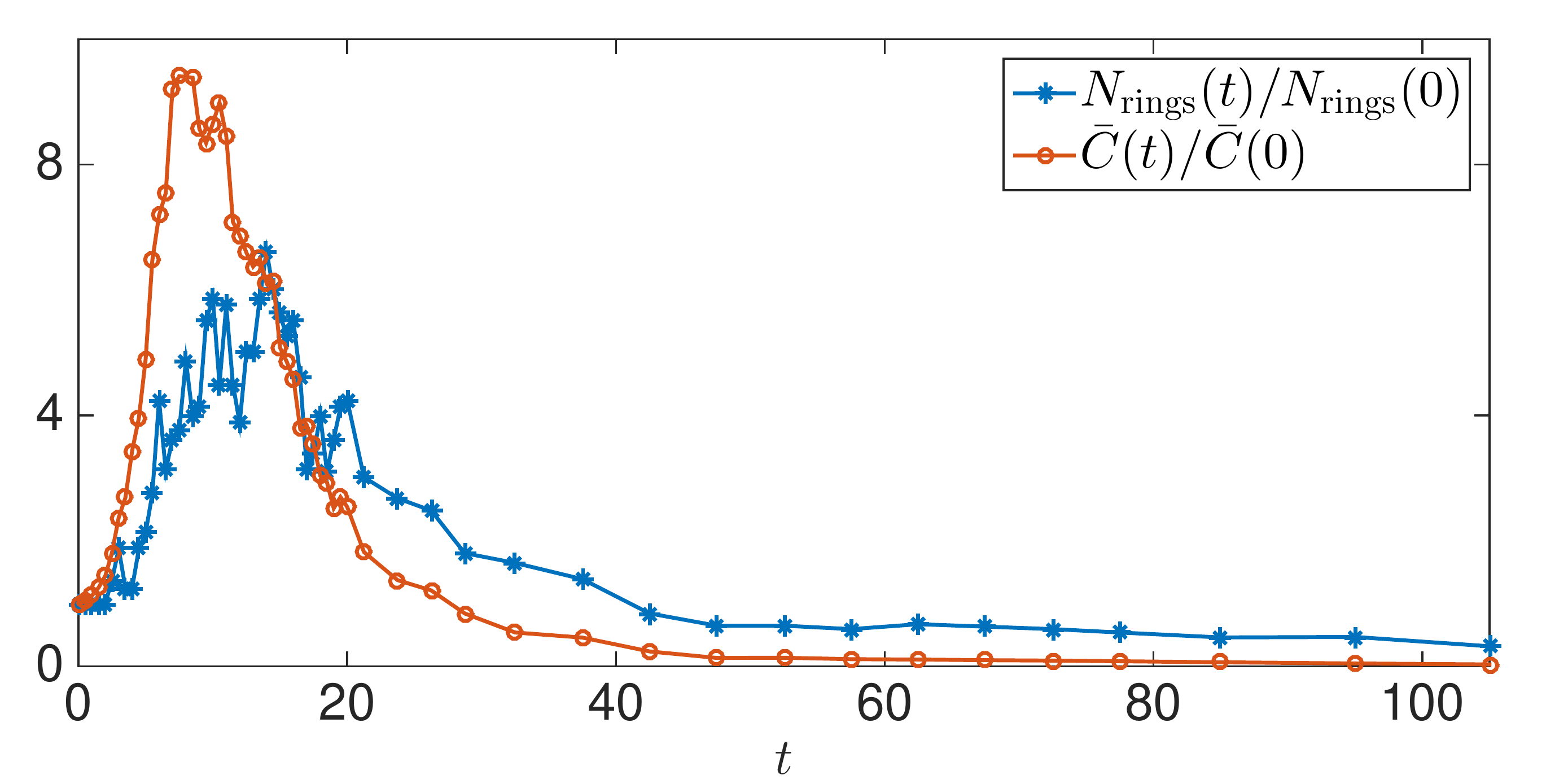}
 \caption{ Temporal evolution of the (normalised) total number of rings and crossing number. At $t=0$, $N_{\rm rings}(0)=128$ and $\bar{C}(0)=758$.}
\label{fig:Complexity}
\end{figure}

In Fig.\ref{fig:Complexity} we plot the evolution of the total number of vortex rings and the average crossing number during the development and decay of the TG tangle. 
Each quantity is normalised with respect to its initial values. As expected from the analysis of the evolution of the total vortex length presented in the Letter, both the number of rings and the average crossing number experience an initial increment. This behaviour can be interpreted as the growth of the complexity in the tangle.

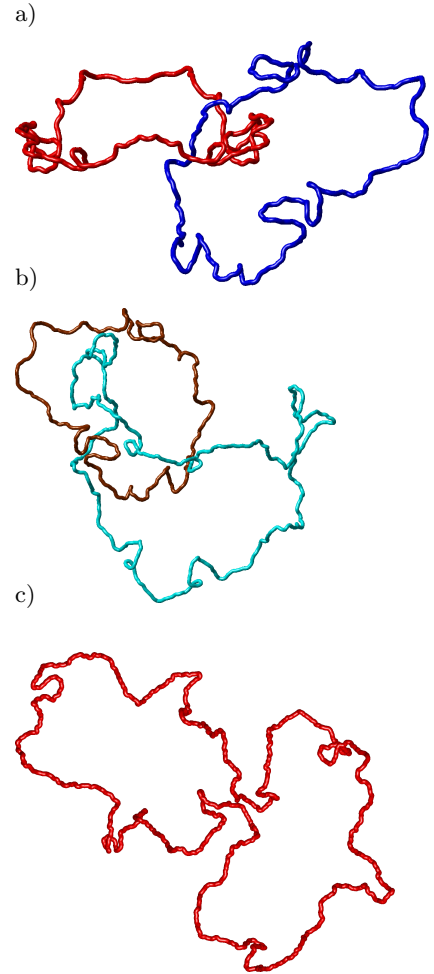
\begin{figure}[h]
 \centering
 \begin{picture}(100,400)
\put(-0,280){\includegraphics[scale=0.5]{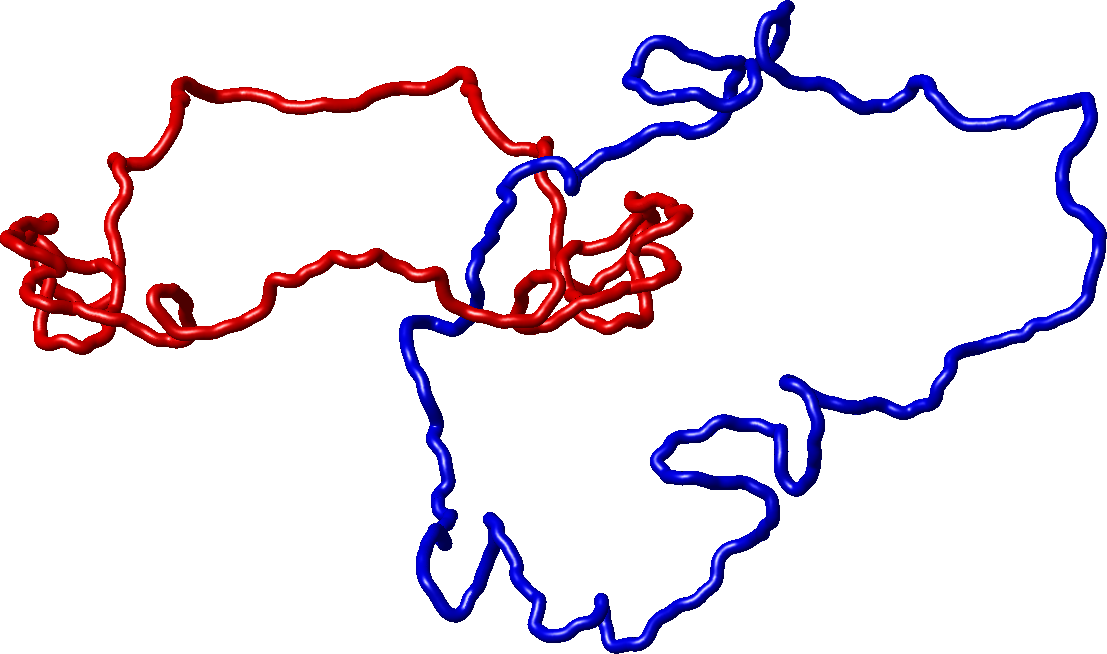}}
\put(-0,160){\includegraphics[scale=0.5]{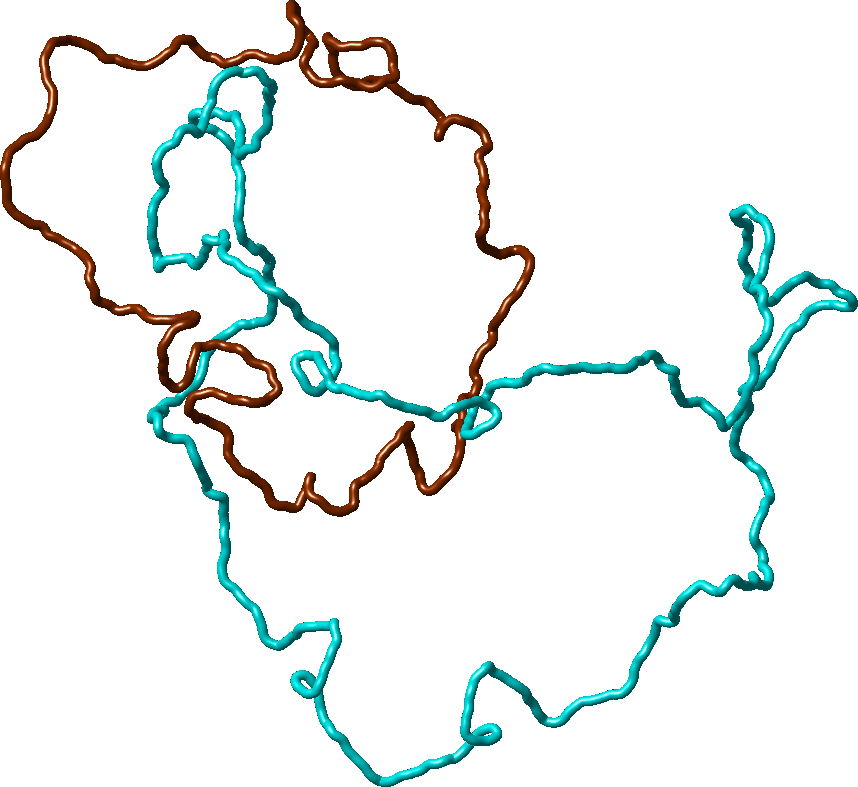}}
\put(-0,20){\includegraphics[scale=0.1]{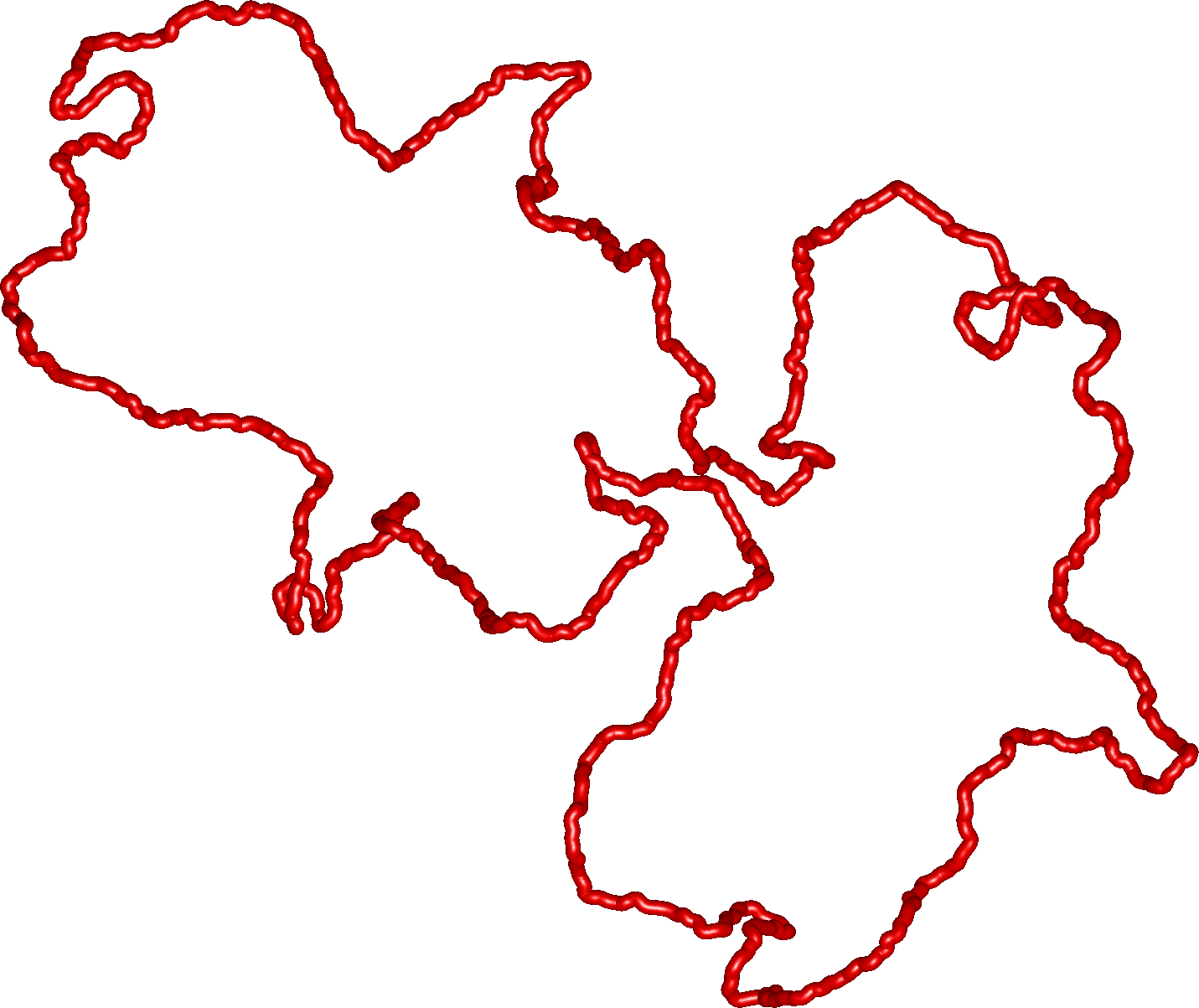}}
\put(-0,380){a)}
\put(-0,280){b)}
\put(0,160){c)}
\end{picture}
 \caption{a) Two linked rings of the Taylor-Green flow found at $t=10$ and b) at time $t=14.5$. c) A ring with high writhe number at $t=24.5$}
\label{fig:Rings}
\end{figure}

As described in the Letter, the linking number takes integer values and provides information of the number of links. 
During the evolution of the Taylor-Green flow, rings can link as presented in Fig.\ref{fig:Rings}a and Fig.\ref{fig:Rings}b. 
These two links were found directly computing the integral for $Lk_{ij}$ for all pair of rings at each time step.
The writhe number takes real values and provide information of self-linking and helical excitations on the filament.
A ring with high value of writhe is displayed in Fig.\ref{fig:Rings}c. Note that the writhe number is not enough to determine whether a ring is self-linked (knotted) or not.
Total link and writhe numbers define the center-line helicity which is a measure of the complexity of the vortex tangle considered.

\section{Appendix D.Kelvin wave spectrum}
In order to obtain the Kelvin wave (KW) spectrum of a ring we proceed as follows. Starting from a ring $R(\sigma)$ provided by the tracking algorithm we obtain a long-scale averaged ring that will be used to determine the amplitude of the KWs. In order to do that, we first compute the natural parametrisation $s$ of the ring and its total length $L$. Then we apply a Gaussian kernel of width $\alpha L$, with $\alpha\in(0,1)$. The obtained convolved ring $R_{\rm smooth}(s)$ is smooth and it can now be used to define the KWs on the ring as
\begin{equation}
R_{\rm KW}(s)=R(s)-R_{\rm smooth}(s).
\end{equation}
By definition, $R_{\rm KW}$ is a periodic set of 3 signals (one for each spatial dimension), hence, the KW spectrum is then defined as
\begin{equation}
n_k=|\widehat{R_{\rm KW}}(k)|^2+|\widehat{R_{\rm KW}}(-k)|^2,
\end{equation}
where $\widehat{R_{\rm KW}}(k)$ is the Fourier transform of $R_{\rm KW}(s)$. 
In \cite{2016arXiv?} we checked that this procedure is able to capture well the KWs superimposed on a ring by performing a series of different tests (different $\alpha$ and amplitude of KWs). In this work we use a value of $\alpha=0.1$, the variation of this fraction only slightly modifies (as expected) the large-scale values of the spectrum, but values in the inertial range remain unchanged.

Note that data from the tracking is not spaced in regular intervals of the natural parametrisation. In order to computed the Fourier transform, the data is re-meshed in a regular partition of $[0,L]$ obtained by a high order interpolation. In reference \cite{2016arXiv?} we have also test this technique by comparing it with the method used in \cite{PhysRevE.86.055301} to detect KWs in a almost straight vortex line and verified that this interpolation does not affect the obtained spectrum.

\bibliographystyle{apsrev4-1}
%\bibliographystyle{aipnum4-1}
%

%\bibliography{textReferences}

\end{document}